# Colloidal Quantum Dot Tandem Solar Cells Using CVD Graphene as An Atomically Thin Intermediate Recombination Layer

Yu Bi[1], Santanu Pradhan[1], Mehmet Zafer Akgul[1], Shuchi Gupta[1], Alexandros Stavrinadis[1], Jianjun Wang[1], Gerasimos Konstantatos[1,2,]*

1. ICFO—Institut de Ciencies Fotoniques, The Barcelona Institute of Science and Technology, Av. Carl Friedrich Gauss, 3, 08860 Castelldefels (Barcelona), Spain

2. ICREA—Institució Catalana de Recerca i Estudis Avançats, Passeig Lluís Companys 23, 08010 Barcelona, Spain

* Corresponding author: Gerasimos.Konstantatos@icfo.eu

ABSTRACT

Two-terminal tandem cell architectures are believed to be an effective way to further improve the power conversion efficiency in solution processed photovoltaics. To design an efficient tandem solar cell, two key issues need to be considered. Firstly, subcells with well-matched currents and complementary absorption characteristics are a prerequisite for high efficiency. Secondly identifying the appropriate intermediate layer (IML) to connect the subcells is necessary to minimize the optical and electronic losses. PbS colloidal quantum dots (CQDs) are a notable choice for the subcells due to their low cost, solution processibility and remarkable wide range band gap tunability. Single layer Graphene (Gr) has been proposed to be a promising IML due to its high transparency and conductivity. Here, as a proof of concept, we demonstrate a solution processed two terminal PbS CQDs tandem solar cell employing chemical vapor deposited Gr as the IML. In doing so, we report a PbS CQD cell comprising subcells with bandgaps of 1.4 and 0.95 eV that delivers power conversion efficiency in excess of 7%, substantially higher than previously reported CQD tandem cells.

With the recent rapid rise in solution processed photovoltaics, power conversion efficiency (PCE) higher than 20% has been reported for perovskite solar cells and values greater than 11% have been achieved in CQD-based solar cells in single junction devices.[1-13] To enter the third generation regime using solution-processed photovoltaics efforts are needed targeting to overcome the single junction S-Q limit. In view of this, tandem solar cells, that stack two or more single junction subcells with different band gaps to harvest photons from the full solar spectrum more efficiently, have attracted growing attention recently[14-20]. To design an efficient tandem solar cell the subcells should possess complementary absorption characteristics and provide well-matched and balanced photocurrent, with minimal electrical losses. A theoretical PCE up to 42% is predicted in a tandem solar cell comprising two current matched subcells with bandgaps of 1.6 eV and 0.95 eV respectively.[21]

Solution processed CQDs stand out as one of the most promising materials for third generation solar cells thanks to the wide bandgap tunabilty (0.7-2.1 eV), particularly pertinent to tandem cells. Significant efforts have been made to optimize high efficiency CQD single junction solar cells,[22-25] and recently remarkable efficiencies of more than 11% have been achieved.[9, 26] However, tandem solar cells based on CQD solar cells has thus far lagged behind, compared with other PV technologies.[17, 27-31] To our knowledge, only few CQD tandem solar cells have been reported.[30-33] The first CQD tandem solar cell was reported by using a ZnO/Au (1nm)/PH neutral PEDOT: PSS interlayer to series connect PbS CQD subcells with optimized band gaps of 1.6eV and 1eV, yielding a relative low $J_{SC}$ of 3.7 mA/cm$^2$, FF of 37%, $V_{OC}$ of 0.91V, and PCE of 1.27%.[30] By introducing an innovative graded recombination layer between two sub cells, Wang et al. significantly improved the performance of PbS CQDs tandem solar cells with a $V_{OC}$ of 1.06 V, $J_{SC}$ of 8.3 mA/cm$^2$, FF of 48% and PCE of 4.2% by developing a multilayer oxide IML involving multiple layer sputtering.[31] In addition, A PCE of 8.9% has been recently reported in a Homo-tandem PbS CQD solar cell by using a thin layer of Au as IML, in which, however, for both subcells the same bandgap CQDs have been employed.[33] A different approach for constructing an IML has also been reported based on a thin tunneling layer of ZnTe/ZnO and demonstrated in a tandem cell comprising PbS CQDs and CdTe NCs. The growth of ZnTe, however involved a sputtering deposition step and the overall performance of the cell was 5%.[32] To date high performance CQD tandem cells have relied on vacuum-deposition based IMLs.[34] To achieve a competitive CQD tandem solar cell technology both in performance and cost, an efficient and vacuum-free-based IML is attractive.

An efficient IML must fulfill the following properties: it has to be optically transparent to avoid parasitic absorption loss, particularly in the infrared; it has to cater for efficient recombination for both electrons and holes to minimize electrical losses; and ideally it should be compatible with roll-to-roll, vacuum-free manufacturing processes to ensure low added manufacturing costs and high throughput. One material that can fulfill concomitantly the afore-mentioned properties is graphene (Gr), in view of its atomically thin and semimetallic nature, high in-plane conductivity and very low and spectrally flat optical absorption. Chemical vapor deposited (CVD) Gr has been proposed as a promising IML in solar cells, [35-38] yet the only demonstration of its potential has been reported in a tandem organic cell, with limited overall performance, likely due to the incompatibility of graphene processing with the environmental requirements of organic solar cells. [37] PbS CQD solar cells have been demonstrated as a more robust material platform towards humidity compared to organic or perovskite solar cells. We have leveraged this feature to develop an efficient tandem CQD solar cell capable of harnessing both the visible and infrared part of solar spectrum using CVD Gr, transferred and deposited in water, as the IML to series connect two PbS CQD subcells. The concept of Gr-CQD tandem solar cell is demonstrated by a high $V_{OC}$ of 1.12 V, a remarkable fill factor of 59%, and a PCE of 8.2% in tandem cells with 885nm and 980 nm PbS CQDs subcells. Moreover we report on a tandem cell that extends also in the short-wave infrared using PbS CQDs with exciton peaks at 865 nm and 1300 nm. This architecture delivers a $V_{OC}$ of approximately 1 V, FF of 50% and a $J_{SC}$ of 14.6 mA/cm$^2$ and a power conversion efficiency of 7.1% significantly higher than the previous record of 4.2% from CQD tandem cells whose energy harvesting extended to the short wave infrared.[31]

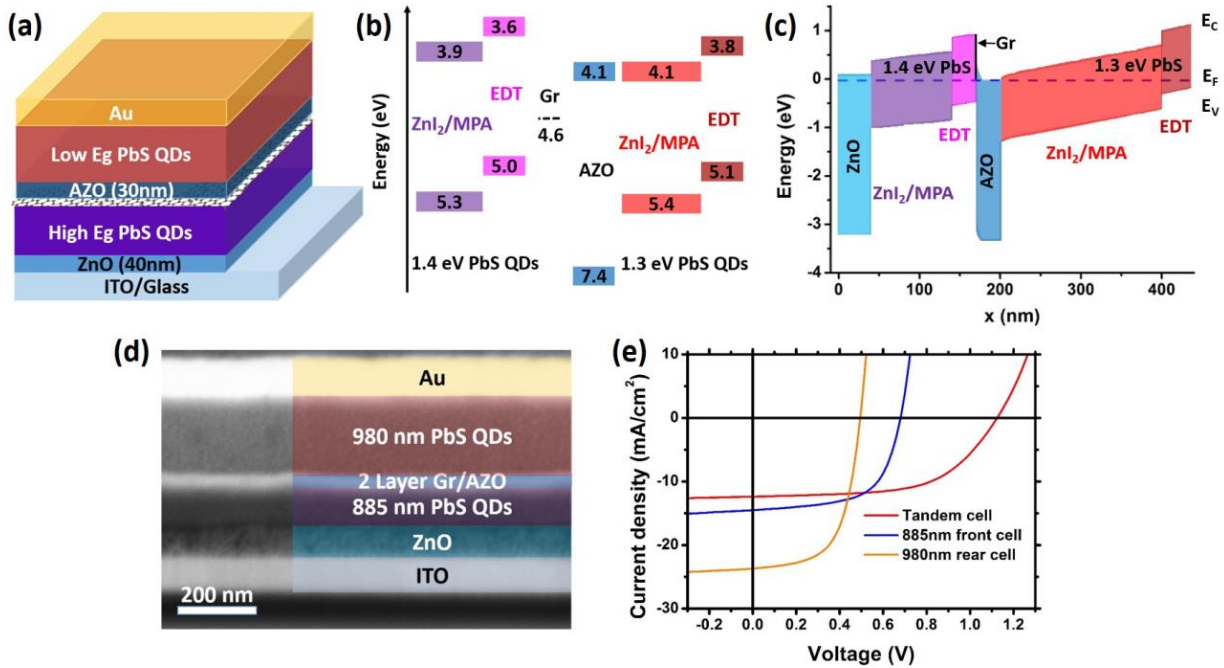

Figure 1. Proof of concept on PbS CQDs tandem solar cells by using Gr as a connecting layer: (a) Schematic structure of the Gr CQDs tandem solar cell consisted of two solution processed PbS CQDs sub cells series connected by CVD Gr, (b) Energy level diagram of the layers comprising the tandem cell, (c) Band diagram of the tandem cell modeled through SCAPS in short-circuit conditions (the input parameters used in the SCAPS software are shown in the Supporting Information on **Table S3**), (d) FIB cross-sectional image of the champion 1.3/1.4 eV CQDs tandem cells, (e) current density voltage characteristics under AM 1.5G illumination of the champion 1.3/1.4 eV tandem cell, and its corresponding dummy front and rear cells.

**Figure 1a** shows the schematic device structure of CQD tandem solar cells developed in this study. Two terminal series-connected tandem cell is designed to harvest the sunlight. Two types of PbS CQDs with complementary absorption range are selected to make front and rear cells, respectively. Front cell comprises of ITO glass substrate, solution processed ZnO nanocrystals (NCs) film and high band gap PbS CQDs film; top cell is made of sputtered alumina doped ZnO (AZO), low band gap PbS CQDs film and thermal evaporated gold film. The IML, which links these two sub cells together, is CVD Gr transferred on top of the front cell by PMMA assisted wet transfer process. As a first proof of concept, a PbS CQD tandem solar cell was fabricated by using CVD Gr to connect the 885 nm PbS CQD front cell and the 980 nm PbS CQD rear cell. The energy levels of the materials comprising the reported tandem cell are illustrated in **Figure 1b** while the band diagram of such a tandem cell in short circuit conditions is shown in **Figure 1c.** The values of the energy levels have been extracted by UPS data reported elsewhere. [25] The work function of Gr at 4.6 eV lies in between the valence band of the electron-blocking layer in the front cell and the electron-accepting layer in rear cell. Previous studies on PbS CQDs/ Gr photodetectors[39], where EDT treated PbS CQD were deposited atop Gr, showed that Gr tended to get p-doped after the CQD deposition, and a favorable charge (hole) transfer takes place from CQDs to graphene. The deep valence band of AZO prevents the hole injection from the front cell, at the same time the electron blocking layer of the front cell prohibits the electron injection from the rear cell. Hence, photogenerated holes from the front cell and electrons from the rear cell recombine efficiently in the IML without any leaking current.

A cross-sectional FIB SEM image of the 885/980nm PbS CQD tandem solar cell is illustrated in **Figure 1d**. The PbS CQD layer in the front cell is well passivated by $ZnI_2$/MPA and EDT as it has been found to be robust enough to survive the water bath during the Gr transfer process. (See **Table S1** in the supporting information.) The bilayer Gr inserted between the subcells is too thin to be seen. In Figure S1 AFM images showed the surface profile before and after single Gr transfer atop PbS CQD layer. The slightly decreased roughness and no visible cracks and winkles indicate the successful Gr transfer atop of PbS CQD film. Atop Gr, a thin layer of AZO (30 nm) is deposited to act as the electron acceptor of the rear cell. The distinguishable layers corresponding to each subcell observed here indicate that Gr layer is strong enough to separate the subcells and protect the front cell from any solvent damage. The optical property of the CVD Gr used in this work is shown in **Figure S2a**. The transmittance of single layer Gr is over 97% as expected in the wavelength range of 400-1500 nm, as for the bilayer Gr, the transparency remains more than 95% in range of 400 -1500 nm. This confirms the high transparency of the CVD Gr we used here. **Figure S2b** presents UV-vis-IR spectra of the front cell (ITO (80 nm)/ZnO (40nm)/ 885 nm PbS CQD (200 nm)), front cell/bilayer Gr and front cell/ bilayer Gr/ AZO (30 nm). The optical loss is consistent with the sum of the transparency of the front cell, bilayer Gr and AZO. In the infrared part beyond 1000 nm, the low transparency originates from the 80 nm ITO substrate and uncorrected reflection and scattering during the measurements. It indicates the huge optical loss in the infrared when applying ITO as IML in tandem cells. However, with Gr as IML, we do not suffer from the problem in this work.

**Table 1**. Champion solar cell performance obtained in the 1.3/1.4eV tandem cell and its corresponding dummy front and rear cells. (The average device performances are quoted in the parentheses).

| Devices | $V_{OC}$ (V) | $J_{SC}$ (mA/cm$^2$) | FF (%) | PCE (%) |
| --- | --- | --- | --- | --- |
| Tandem cell | 1.12 | 12.38 | 59 | 8.20 |
| | (1.05±0.04) | (11.32±1.03) | (57±3) | (6.79±0.89) |
| Front cell (885 nm PbS) | 0.68 | 14.50 | 62 | 6.08 |
| | (0.66±0.02) | (14.32±1.26) | (59±3) | (5.52±0.41) |
| Rear cell (980 nm PbS) | 0.5 | 23.68 | 60 | 7.06 |
| | (0.46±0.02) | (22.99±0.43) | (54±3) | (5.77±0.54) |

We plot the current density-voltage (J-V) curves of the champion CQDs tandem solar cell, and its corresponding dummy front and rear subcells in **Figure 1e**. The dummy front cell made of ITO (80 nm)/ZnO (40 nm)/885 nm PbS CQDs (100 nm)/Au, yields a $V_{OC}$ of 0.68V, a short circuit current ($J_{SC}$) of 14.5 mA/cm$^2$, and a fill factor (FF) of 62%, resulting in a PCE of 6.08% (**Table 1**) when operated as a single junction cell. The dummy rear cell with device structure of ITO (80 nm)/AZO (30 nm)/980 nm PbS (200 nm)/Au yields a $V_{OC}$ of 0.5 V, $J_{SC}$ of 23.68 mA/cm$^2$, a FF of 60% and a PCE of 7.06 %. The individual subcells are fabricated with the same parameters as in the tandem cells. As expected, the device performance of the tandem solar cell outperforms that of the individual subcells. The value of the $V_{OC}$ (1.12V) attained in the champion tandem cell is very close to the sum of the $V_{OC}$ (1.18V) observed from two single subcells in series and it indicates good Ohmic contact between the photoactive layers and the CVD Gr interface. The $J_{SC}$ of 12.38 mA/cm$^2$ recorded in the tandem solar cell, is very close to the $J_{SC}$ of 14.5 mA/cm$^2$ observed in the front subcell. The above confirm that CVD bilayer graphene acts as an effective IML that allows the photogenerated electrons and holes from the adjacent junctions to meet and recombine efficiently with minimal optical and electronic losses.

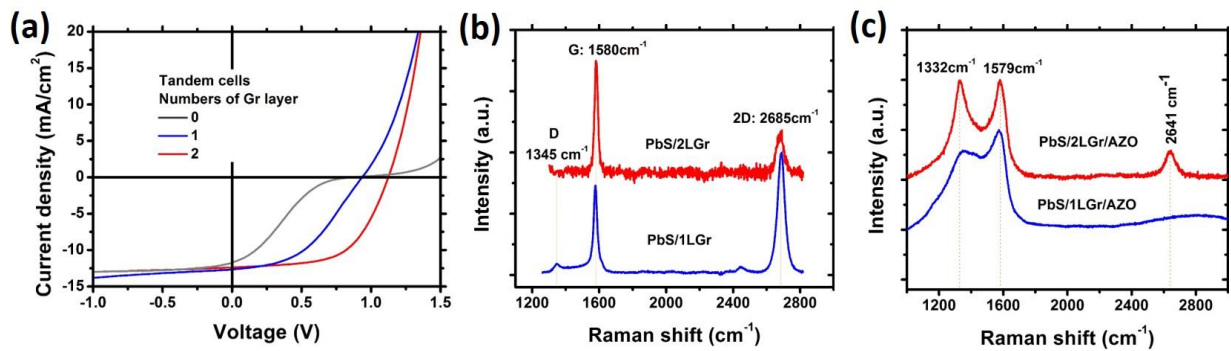

Figure 2. The effect of Gr on PbS tandem solar cells. (a). JV characterization of tandem devices with different layers of Gr, Raman spectra of single and bilayer Gr on top of PbS CQD films (b) and sandwiched between PbS CQD film and AZO (c).

**Table 2.** The effect of Gr layer numbers on PbS CQDs tandem cells device performance, both the champion and average (quoted in the parentheses) device performances are included.

| Gr layers | $V_{OC}$ (V) | $J_{SC}$ (mA/cm$^2$) | FF (%) | PCE (%) |
|---|---|---|---|---|
| 0 | 0.86 (0.81±0.06) | 11.71 (9.53±1.25) | 22 (15±4) | 2.24 (1.17±0.47) |
| 1 | 0.94 (0.95±0.01) | 12.64 (11.70±0.50) | 44 (42±1) | 5.22 (4.67±0.27) |
| 2 | 1.12 (1.05±0.04) | 12.38 (11.32±1.03) | 59 (57±3) | 8.20 (6.79±0.89) |

To account for the effect of Gr in the tandem solar cells, we fabricated tandem devices with and without Gr as an IML. The beneficial role of Gr as IML in CQDs tandem solar cells was experimentally demonstrated. J-V curves of the tandem solar cells with different number of Gr layers as IML are shown in **Figure 2a**. By introducing Gr into the CQDs tandem solar cells, FF is significantly improved from 9% up to 59% in **Table 2**. Without Gr, the s-shape appears in the JV curve, yet the s-shape of the J-V curve is strongly suppressed by introducing Gr. With bilayer Gr, the s-shape completely disappeared in the JV curve. The ohmic-like contact forms by inserting CVD Gr between AZO and ITO evidenced by JV measurements in **Figure S3.** We sought to identify the origin of the optimized performance of bilayer versus single layer Gr in the cells and we have used Raman spectroscopy to assess the quality of the Gr in the two cases. [40-41] **Figure 2b, c** shows the Raman spectra of the Gr before and after AZO deposition. The two most prominent peaks, G band (1580 cm$^{-1}$) and 2D band (2685 cm$^{-1}$) are observed as expected on the Gr/front cell samples in **Figure 2b**. The intensity ratio of 2D and G peak ($I_{2D}/I_G$) can be used to estimate the number of layers and the doping level of Gr.[42] $I_{2D}/I_G$ of single and bilayer Gr atop PbS CQD layer is 2.5 and 0.71, respectively. This confirms that no obvious defects are formed and the quality of Gr remains high upon transferring onto the front cell.[43] Then we sought to assess the effect of AZO sputtering atop the Gr as employed in our devices, as sputtering is known to damage the Gr layer. [44] **Figure 2c** shows that the defect-related D band around 1330 cm$^{-1}$ appears for both single and bilayer of Gr after AZO deposition, although the G band peak at 1580 cm$^{-1}$ is not affected by the deposition. For the single layer Gr, the 2D band around 2685 cm$^{-1}$ disappears after AZO deposition. Since 2D band is the most prominent feature in Gr, it is always seen even when no D peak is present. [42, 45] This points to single layer Gr deformation upon AZO deposition. The appearance of the D and G band in the Raman spectra most likely point to the existence of Gr oxide.[46-47] The oxidation of the single layer Gr during AZO deposition is possible since oxygen (O$_2$/Ar=1/10) is employed during the deposition process. In the case of bilayer Gr, the 2D band blue shifts to 2641 cm$^{-1}$, which indicates the presence of single layer Gr. Thus in the case of bilayer Gr after AZO deposition, the top-layer exposed to AZO transforms to Gr oxide, leaving the bottom Gr layer intact. Therefore, the use of bilayer Gr, to begin with, has been instrumental in achieving a high performance IML since the top layer is damaged by the AZO sputtering process and acts as a sacrificial layer to ensure the presence of at least one Gr layer in the tandem device that acts as the recombination layer.

Leveraging the most promising advantage in PbS CQD based solar cells, which is the wide range band gap tunability, allows us to select close-to-ideal band gap combinations for the tandem cells to harvest as much solar light as possible. So far, the most popular CQDs solar cells reported in the literature is based on PbS CQDs with first exaction peak around 900 nm. With our recent efforts in developing highly efficient PbS solar cells, we have achieved a very high $V_{OC}$ in CQD single junction solar cells with PbS CQD bandgap around 1.3 eV of 0.65 V, [25, 48] and a high $J_{SC}$ of 34 mA/cm$^2$ with the 0.9 eV bandgap PbS CQD devices (see supporting information **Table S2**).[49] Here we have extended our Gr-CQD tandem solar cell concept to connect these two single cells in series. For the series connected tandem solar cells it is critical to balance the current generated in the front and rear subcells to deliver optimum performance, since the $J_{SC}$ in the tandem cell strongly depends on the current in the subcells. To do so, we first employed optical modeling based on the transfer matrix method to provide guidance for the tandem device fabrication.[50-52]

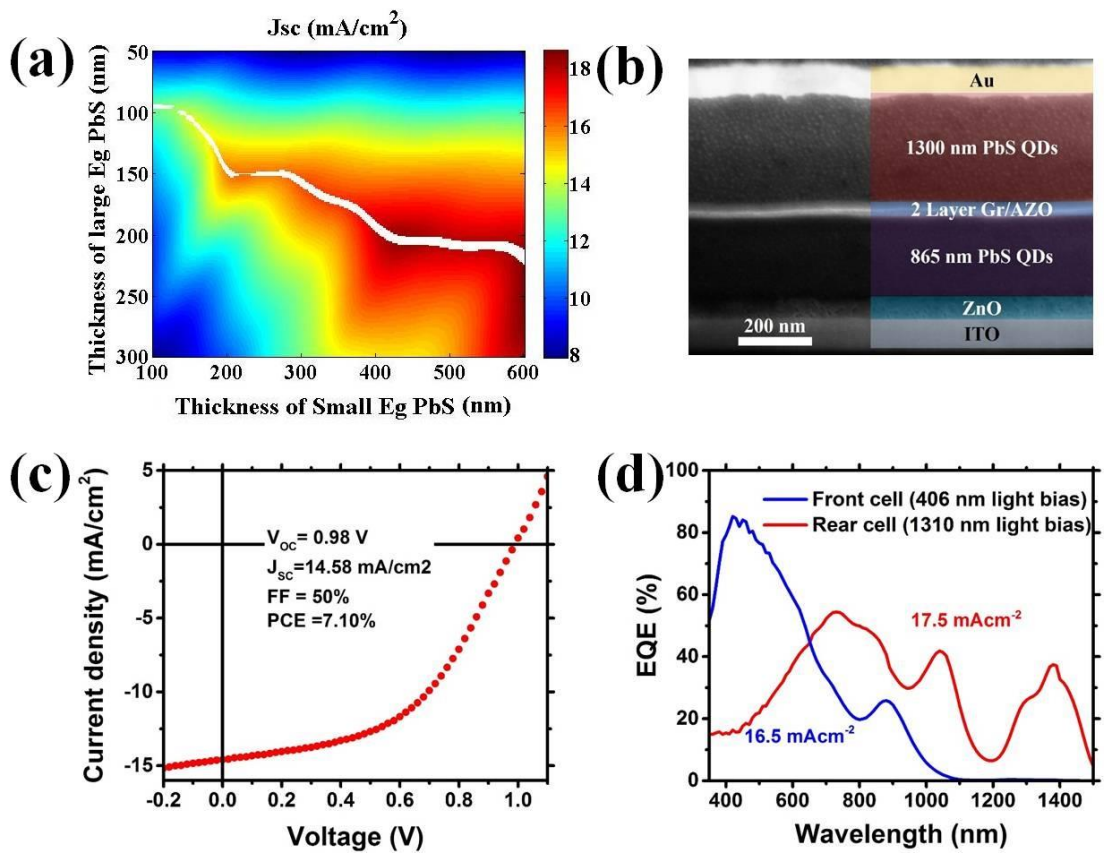

Figure 3. (a) Simulated short-circuit current density generated in the tandem device as a function of the active layer thicknesses. The white trace shows the current matched active layer thicknesses, (b) FIB cross-sectional image of the CQD tandem solar cell, (c) JV characterization under AM 1.5G illumination of the champion tandem cell and (d) external quantum efficiency spectra for the subcells, measured with laser bias at 406 nm and 1310 nm respectively.

**Figure 3a** shows the simulated $J_{SC}$ extracted from the tandem device as a function of the thickness of the active layers in the front and rear cells. The refractive index n and extinction coefficient k of the different layers in tandem cell are calculated based on the spectroscopic ellipsometer measurements. (See **Figure S4** in the supporting information) The simulation shows that an optimum value of $J_{SC}$ above 18 mA/cm$^2$ can be reached when the thicknesses of the optimized 865nm PbS CQD layer and 1300 nm PbS CQD layer are around 200 nm and 435 nm, respectively. **Figure 3b** presents the FIB cross sectional SEM image of the tandem devices based on 865 nm and 1300 nm PbS CQDs subcells. Bilayer Gr was introduced as the IML to connect the subcells. Similar to Figure 1b, the distinguishable layers corresponding to each subcell can be clearly observed in the **Figure 3b** (bilayer Gr inserted between the subcells is too thin to be resolved). The JV curve and its key photovoltaic parameters of the champion tandem cell is presented in **Figure 3c**. A $V_{OC}$ of 0.98 V is measured, which is very close to the sum of the individual subcells (1.08 V). The $J_{SC}$ of the tandem cell is 14.58 mA/cm$^2$, and the FF is 50%, leading to an overall performance of 7.1%. The simulated $J_{SC}$ of this specific tandem cell is 14.4 mA/cm$^2$ corresponding to the active layer thicknesses of 210 nm and 290 nm for the front and rear cell in Figure 3b, in good agreement with the measured tandem $J_{SC}$ of 14.58 mA/cm$^2$ under AM1.5 illumination. To validate the optical modeling we used in Figure 3a, several tandem devices with varied active layer thickness in both front and rear cell were made. As shown in Table S4 and S5 in the supporting information the measured $J_{SC}$ trend is consistent with the values from the optical modeling of Figure 3a. The overall champion PCE of 7.1%, in this case, is relatively lower than the one of 8.2% in the mismatched bandgap tandem cell presented in **Figure 1**, due to the slightly lower performance of the rear SWIR subcell in terms of $V_{OC}$ and FF compared to the NIR cells used in Figure 1. To further account for the $J_{SC}$ attained in the tandem cell, the EQE spectra of the front and rear subcells in the tandem cell were measured by using light bias from 1310 nm and 406 nm lasers, respectively (see **Figure S6** in the supporting information).[53] As shown in **Figure 3d**, the front cell absorbs most of the high energy photons within the range of 300-700 nm, and the maximum response over 80 % is reached from 400 to 480 nm, this is also consistent with the transmission spectra of the front cell (**Figure S2**). The rear cell absorbs most of the low energy photons within the range of 600-1500 nm. The front cell shows around 20% EQE in the first exciton peak range, and the rear cell still shows 15- 40 % EQE in the range of 350-700 nm due to the absorption overlap in these ranges. The integrated $J_{SC}$ from the EQE are 16.5 mA/cm$^2$ for the front cell and 17.5 mA/cm$^2$ for the rear cell, demonstrating that the two subcells are fairly well matched.

In summary, we have demonstrated that an atomically thin layer of graphene acts as a very efficient recombination layer for CQD tandem solar cells, eliminating the need for complex multi-layer sputtering processes involved in prior reports. In doing so, we report a power conversion efficiency of 8.2% from a tandem architecture involving slightly mismatched bandgaps of single junction PbS QD cells. By leveraging the quantum size effect we have also reported a graphene-IML-based tandem cell comprising two PbS QD cells with complementary bandgaps of 1.4 eV and 0.95 eV reaching a power conversion efficiency in excess of 7%, the highest reported efficiency for a CQD tandem cell. Future efforts in optimizing further the performance of the subcells and combining this with perovskite and other PV technologies are foreseen to realize high performance two-terminal solution-processed tandem photovoltaics.

**Acknowledgements**

We acknowledge financial support from the European Research Council (ERC) under the European Union's Horizon 2020 research and innovation programme (grant agreement No 725165), the Spanish Ministry of Economy and Competitiveness (MINECO) and the "Fondo Europeo de Desarrollo Regional" (FEDER) through grant MAT2014-56210-R. This work was also supported by AGAUR under the SGR grant (2014SGR1548) and by European Union H2020 Programme under grant agreement n°696656 Graphene Flagship. We also acknowledge financial support from Fundacio Privada Cellex, the CERCA Programme and the Spanish Ministry of Economy and Competitiveness, through the "Severo Ochoa" Programme for Centres of Excellence in R&D (SEV-2015-0522). We also thank Graphenea S.A. for providing us the CVD graphene, and we thank Dr. Teresa Galan and Gabriele Navickaite for the training of graphene transfer. We thank Dr. Teresa Galan for the AFM measurements. We are also thankful Dr. Iñigo Ramiro for fruitful discussions about EQE measurements.

**Competing financial interests**

The authors declare no competing financial interests.

**Supporting Information**

Experimental section, AFM images of the PbS QDs film before and after single layer Gr transfer atop, UV-vis-IR spectra, additional device performance tables, the refractive index and extinction coefficient values from the ellipsometer measurements, the simulated field distribution, the simulated Jsc generated in the tandem devices as a function of the active layer with different intermediate recombination layers, SCAPS simulation parameters